# Analytical Modeling of *I-V* characteristics using 2D Poisson Equations in AlN/*β*-Ga$_2$O$_3$ HEMT

[1]R. Singh, [1]T. R. Lenka, [2]D. K. Panda, [3]H. P. T. Nguyen, [4]N. El. I. Boukortt, [5]G. Crupi

[1]Department of Electronics and Communication Engineering, National Institute of Technology Silchar Assam, India (Email: rajan_rs@ece.nits.ac.in , trlenka@ieee.org)
[2]School of Electronics, VIT-AP University, Amaravati, AP, 522237 India (deepak.panda@vitap.ac.in)
[3]Department of Electrical and Computer Engineering, New Jersey Institute of Technology, Newark, NJ 07102, USA (E-mail: hieu.p.nguyen@njit.edu )
[4]Electronics and Communication Engineering Department, Kuwait College of Science and Technology, Doha, Kuwait (n.boukortt@kcst.edu.kw)
[5]BIOMORF Department, University of Messina, Messina 98125, Italy (E-mail: crupig@unime.it)

***Abstract*:** In this paper, physics-based analytical models using two-dimensional (2D) Poisson equations for surface potential, channel potential, electric field, and drain current in AlN/β-Ga$_2$O$_3$ high electron mobility transistor (HEMT) are presented. The analytical expression of different quantities is achieved based on full depletion approximation of the AlN barrier layer and polarization charge induced unified two-dimensional electron gas (2DEG) charge density model. For the validation of the developed model, results are compared with 2D numerical simulation results and a good consistency is found between the two. The drain current model is also validated with experimental results of a similar dimension device. The developed model can be a good reference for different β-Ga$_2$O$_3$-based HEMTs.

***Keywords*:** 2DEG, AlN/*β*-Ga$_2$O$_3$, HEMT, Poisson equation, Surface potential

## 1. Introduction

Nowadays, Gallium-oxide (Ga$_2$O$_3$) is being explored for its possible application in the decade-old high electron mobility transistors (HEMTs) technology, currently dominated by III-nitride material like gallium-nitride (GaN). GaN HEMTs have shown excellent performance for high power and high-frequency applications on the back of exciting features such as higher two-dimensional electron gas (2DEG) density $n_s$ ~ $10^{13}$ cm$^{-2}$, very high 2DEG mobility ~ 1500 cm$^2$/Vs. However, rising emerging applications like electric vehicles (EVs) and robotics have expanded the investigation horizon from wide bandgap (WBG) to ultra-wide bandgap (UWB) semiconductors. Among the UWB semiconductors, β-Ga$_2$O$_3$ is found to be more suitable for high voltage applications due to its ultra-large bandgap of ~ 4.9 eV, and breakdown field of ~ 8 MV. Moreover, β-Ga$_2$O$_3$ single crystal substrate can be grown using melt-based techniques, which ensures large size and affordable wafers for different device technologies.

To date, various β-Ga$_2$O$_3$ based experimental devices like Schottky diodes [1], [2], metal-oxide-semiconductor field-effect transistors MOSFETs[3]–[6], and β-(Al$_x$Ga$_{1-x}$)$_2$O$_3$/Ga$_2$O$_3$ (AGO/GO) HEMTs [7], [8] with excellent DC and RF performances have been reported. However, since β-Ga$_2$O$_3$ does not has any polarization property, and small band offset in AGO/GO HEMTs limit their performance as compared to their AlGaN/GaN counterpart. On the other hand, epitaxial layers of III-nitrides (AlN, GaN, InN) have been grown on β-Ga$_2$O$_3$ and potential AlN/β-Ga$_2$O$_3$ HEMT applications are anticipated [9]–[12]. Moreover, one of our previous works in AlN/β-Ga$_2$O$_3$ HEMT estimated exceptional RF performance—cut-off frequency $f_T$ of 167 GHz, and output power $P_{OUT}$ of 2.91 W/mm [13]. Since the 2DEG density $n_s$ in HEMTs has a critical role in device operation, recently we have reported validation of estimated $n_s$ in AlN/β-Ga$_2$O$_3$ HEMT through analytical modeling in [14]. Various physics-based models for different parameters in AlGaN/GaN HEMTs have been reported [15]–[20], but almost none for β-Ga$_2$O$_3$-based HEMTs. Looking at the accuracy of the mathematical models over empirical and approximation models, device physics-based mathematical models for β-Ga$_2$O$_3$ based HEMT will be useful to analyze and validate the estimated results. Here, we present physics-based mathematical models for surface potential and electric fields and extended them to get the current-voltage model in AlN/β-Ga$_2$O$_3$ HEMT with an etched AlN barrier layer. For the validation of results, developed model results are compared with simulation as well as experimental results earlier published for similar β-Ga$_2$O$_3$ devices.

## 2. Device Structure

The cross-sectional schematic of the analysed device is shown in Fig. 1. The epi-layer sequence is as follows: on a semi-insulating β-Ga$_2$O$_3$ substrate an n-type doped 100 nm β-Ga$_2$O$_3$ buffer layer with $N_D = 1.5 \times 10^{16}$ /cm$^3$ exists, followed by 10 nm AlN barrier layer which is undoped. Source and drain contacts are set as Ohmic with contact resistance of 0.4 Ω-mm. The gate contact is assumed as Schottky and a barrier potential of 0.8 eV is set. The device dimensions are kept the same as [5] with gate length $L_G$ of 2.8 μm, gate-drain spacing $L_{GD}$ of 1.3 μm, and gate-source spacing $L_{GS}$ of 0.4 μm. The gate width W is equal to 100 μm, and the output drain current is normalized with this quantity. In the simulation framework different physics-based models like Shockley-Read-Hall (SRH) recombination, Fermi-Dirac for carrier statistics, and negative differential conductivity (NDC) to capture electron velocity saturation effect are evoked. Default material parameters and spontaneous as well as piezoelectric polarization models are used from [21] for AlN barrier material. The β-Ga$_2$O$_3$ material parameters used in the simulations and model development are taken from [22], [23], Table 1.

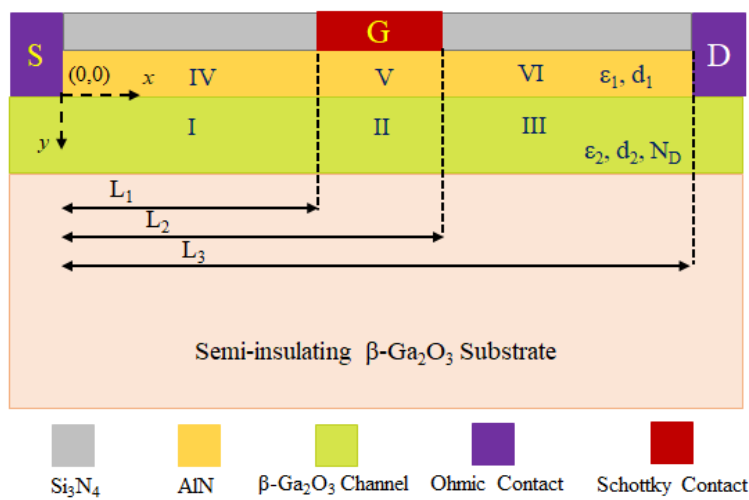

Fig. 1: 2D cross-sectional view of AlN/β-Ga$_2$O$_3$ HEMT.

Table 1. Material parameters for β-Ga$_2$O$_3$ [22], [23].

| Symbol | Quantity | Values | Symbol | Quantity | Values |
|---|---|---|---|---|---|
| $E_g$ | Energy bandgap | 4.9 eV | $N_V$ | Valence band density | $2.86 \times 10^{18}$ |

| $\chi$ | Electron affinity | 3.15 eV | $\varepsilon_r$ | Relative permittivity | 10 |
|---|---|---|---|---|---|
| $m_e^*$ | Electron effective mass | $0.25 m_0$ | $a_0$ | Lattice constant | 3.04 |
| $D$ | Density of states | $1.045 \times 10^{18}$ m$^{-2}$ V$^{-1}$ | $\mu_{n0}$ | Electron mobility | 140 cm$^2$ /Vs |
| $N_C$ | Conduction band density | $3.13 \times 10^{18}$ | $\mu_{p0}$ | Hole mobility | 50 cm$^2$ /Vs |

A list of symbols used in the model development is given in Table 2.

Table 2. Symbols used in analytical modeling

| Symbol | Quantity | Symbol | Quantity |
|---|---|---|---|
| $\varepsilon_1, \varepsilon_2$ | Permittivity of AlN and β-Ga$_2$O$_3$ material respectively | q | Unit electron charge |
| d$_1$, d$_2$ | Thickness of AlN and β-Ga$_2$O$_3$ layer respectively | V$_{FB}$ | Flat band voltage |
| N$_D$ | Channel doping concentration | V$_{bi}$ | Built-in potential |
| Φ$_m$ | Gate metal workfunction | E$_F$ | Fermi potential |
| n$_s$ | 2DEG density | k | Boltzmann constant |
| V$_{OFF}$ | Threshold voltage | ΔE$_C$ | Conduction band offset |
| V$_{th}$ | Thermal voltage | χ | Electron affinity |
| C$_G$ | Gate capacitance | T | Temperature |

A physics-based analytical model for surface potential, electric field, and drain current are developed by solving two-dimensional (2-D) Poisson's equation in the channel region marked as Region I, II, and III as shown in Fig. 1. The 2-D Poisson's equation for the potential function φ (x, y) in the channel (0 ≤ $x$ ≥ L$_3$) is given by:

$$\frac{d^2\varphi_i(x,y)}{dx^2} + \frac{d^2\varphi_i(x,y)}{dy^2} = -\frac{qN_D}{\varepsilon_2}, \quad i = 1, 2, 3 \quad (1)$$

$$\frac{d^2\varphi_i(x,y)}{dx^2} + \frac{d^2\varphi_i(x,y)}{dy^2} = 0, \quad i = 4, 5, 6 \quad (2)$$

where N$_D$, and ε$_2$ are doping concentration and permittivity of β-Ga$_2$O$_3$ buffer layer. Since potential at the AlN/β-Ga$_2$O$_3$ interface is continuous, it is only solved for the region I, II, and III.

The potential function φ(x, y) along the y-direction in the channel can be approximated by a parabolic expression such as:

$$\varphi_i(x,y) = \varphi_{si}(x) + A_i(x)y + A_{i+1}(x)y^2, for\ i = 1, 2, 3 \quad (3)$$

where $\varphi_{si}(x)$ are the surface potential, and $A_i(x)$ are arbitrary constants for the region I, II, and III that are to be determined using boundary conditions. The Poisson's equation (1) is solved simultaneously for all three regions in the channel using solution expressed in (3), and given as:

$$\varphi_1(x,y) = \varphi_{s1}(x) + A_1(x)y + A_2(x)y^2, for\ 0 \leq x \leq L_1 \quad (4)$$

$$\varphi_2(x,y) = \varphi_{s2}(x) + A_3(x)y + A_4(x)y^2, for\ L_1 \leq x \leq L_2 \quad (5)$$

$$\varphi_3(x,y) = \varphi_{s3}(x) + A_5(x)y + A_6(x)y^2, for\ L_2 \leq x \leq L_3 \quad (6)$$

where $\varphi_1(x,y)$, $\varphi_2(x,y)$, $\varphi_3(x,y)$ and $\varphi_{s1}(x)$, $\varphi_{s2}(x)$, $\varphi_{s3}(x)$ are the channel potential and surface potential in the channel in the region I, II, and III respectively. Furthermore, above equations (4), (5), and (6) are solved using the following boundary conditions for $\varphi_i(x,y)$ are given by:

$$\left.\frac{d\varphi_1(x,y)}{dy}\right|_{y=0} = \frac{\varepsilon_1}{\varepsilon_2}\frac{\varphi_{s1}(x) - V'_{GS}}{d_1} \tag{7}$$

$$\left.\frac{d\varphi_1(x,y)}{dy}\right|_{y=-d_1} = \frac{-qn_{s1}}{\varepsilon_1} = -E_1 \tag{8}$$

$$\left.\frac{d\varphi_2(x,y)}{dy}\right|_{y=0} = \frac{\varepsilon_1}{\varepsilon_2}\frac{\varphi_{s2}(x) - V'_{GS}}{d_1} \tag{9}$$

$$\left.\frac{d\varphi_2(x,y)}{dy}\right|_{y=-d_1} = \frac{-qn_{s2}}{\varepsilon_1} = -E_2 \tag{10}$$

$$\left.\frac{d\varphi_3(x,y)}{dy}\right|_{y=0} = \frac{\varepsilon_1}{\varepsilon_2}\frac{\varphi_{s3}(x) - V'_{GS}}{d_1} \tag{11}$$

$$\left.\frac{d\varphi_3(x,y)}{dy}\right|_{y=-d_1} = \frac{-qn_{s1}}{\varepsilon_1} = -E_3 \tag{12}$$

$$\varphi_{s1}(x)|_{x=0} = V_{bi} \tag{13}$$

$$\varphi_{s1}(x)|_{x=L_3} = V_{DS} + V_{bi} \tag{14}$$

$$V'_{GS} = V_{GS} - V_{FB} \tag{15}$$

where $V_{FB} = \Phi_m - \Phi_{s1}$, where $\Phi_{s1} = \chi_1 + E_{G1}/2$, since the AlN barrier layer is undoped.

Since β-Ga$_2$O$_3$ does not possess any polarization property, the total polarization charge of the AlN/ β-Ga$_2$O$_3$ heterostructure is a sum of spontaneous as well as piezoelectric polarization of the AlN barrier layer. Total polarization of the AlN barrier $P_T = P_{sp} + P_{pi}$, where spontaneous polarization $P_{sp} = -0.09$ C/m$^2$ [21], and piezoelectric polarization $P_{pi}$ is calculated as given in [21]:

$$P_{pi} = 2\left(\frac{a_s - a_0}{a_0}\right)e_{31} - \frac{C_{13}}{C_{33}}e_{33} \tag{16}$$

where $a_s$, $a_0$ are lattice constants of AlN and β-Ga$_2$O$_3$; $e_{31}$, $e_{33}$ and $C_{13}$, $C_{33}$ are piezoelectric and elastic constants for AlN material respectively.

Based on full depletion of the AlN barrier layer, the polarization induced 2DEG charge density is given by [17]

$$n_s = \frac{\varepsilon_1}{qd_1}(V_{GO} - E_F - \varphi_i(x)) \tag{17}$$

where $V_{GO} = V_{GS} - V_{OFF}$, and $\varphi_i(x)$ is the channel potential in regions I, II, III for $i = 1, 2, 3$. Furthermore, cut-off voltage $V_{OFF}$ is given as

$$V_{OFF} = \Phi_B - \Delta E_C - \frac{P_T}{C_G} \tag{18}$$

Where $\Phi_B$ is the Schottky barrier height; $\Delta E_C$ is the conduction band offset between AlN and β-Ga$_2$O$_3$ at the heterointerface; $P_T$ is the total polarization charge density, and $C_G$ is the gate capacitance.

Since surface potential and the lateral electric field between the two consecutive regions at the heterointerface are continuous. The associated boundary conditions at the interface of region I-II and region II-III are given as

$$\varphi_{s1}(x)|_{x=L_1} = \varphi_{s2}(x)|_{x=L_2} \tag{19}$$

$$\frac{d\varphi_{s1}(x)}{dx}\bigg|_{x=L_1} = \frac{d\varphi_{s2}(x)}{dx}\bigg|_{x=L_1} \tag{20}$$

$$\varphi_{s2}(x)|_{x=L_2} = \varphi_{s3}(x)|_{x=L_2} \tag{21}$$

$$\frac{d\varphi_{s2}(x)}{dx}\bigg|_{x=L_2} = \frac{d\varphi_{s2}(x)}{dx}\bigg|_{x=L_2} \tag{22}$$

Now using boundary conditions of (7) to (12), all six constants of (4) to (6) are determined to get the AlN/β-Ga$_2$O$_3$ HEMT channel potential as

$$\varphi_1(x,y) = \varphi_{s1}(x) + \frac{\varepsilon_1}{\varepsilon_2}\left(\frac{\varphi_{s1}(x) - V'_{GS}}{2d_1}\right)y + \left\{\frac{-E_1}{2d_1} + \frac{\varepsilon_1}{\varepsilon_2}\left(\frac{\varphi_{s1}(x) - V'_{GS}}{2d_1}\right)\right\}y^2 \tag{23}$$

$$\varphi_2(x,y) = \varphi_{s2}(x) + \frac{\varepsilon_1}{\varepsilon_2}\left(\frac{\varphi_{s2}(x) - V'_{GS}}{2d_1}\right)y + \left\{\frac{-E_2}{2d_1} + \frac{\varepsilon_1}{\varepsilon_2}\left(\frac{\varphi_{s1}(x) - V'_{GS}}{2d_1}\right)\right\}y^2 \tag{24}$$

$$\varphi_3(x,y) = \varphi_{s3}(x) + \frac{\varepsilon_1}{\varepsilon_2}\left(\frac{\varphi_{s3}(x) - V'_{GS}}{2d_1}\right)y + \left\{\frac{-E_3}{2d_1} + \frac{\varepsilon_1}{\varepsilon_2}\left(\frac{\varphi_{s1}(x) - V'_{GS}}{2d_1}\right)\right\}y^2 \tag{25}$$

Now substituting values of the channel potential obtained above in (1), new equations for the region I, II, and III are obtained as

For region I ($0 \leq x \leq L_1$)

$$\frac{d^2\varphi_{s1}(x)}{dx^2} - \frac{\varepsilon_1}{\varepsilon_2 d_1^2}\varphi_{s1}(x) = \frac{\varepsilon_1}{\varepsilon_2 d_1^2}\left\{\frac{-qN_D}{\varepsilon_2}\frac{\varepsilon_2 d_1^2}{\varepsilon_1} + \frac{E_1}{d_1}\frac{\varepsilon_2 d_1^2}{\varepsilon_1} - V'_{GS}\right\} \tag{26}$$

For region II ($L_1 \leq x \leq L_2$)

$$\frac{d^2\varphi_{s2}(x)}{dx^2} - \frac{\varepsilon_1}{\varepsilon_2 d_1^2}\varphi_{s2}(x) = \frac{\varepsilon_1}{\varepsilon_2 d_1^2}\left\{\frac{-qN_D}{\varepsilon_2}\frac{\varepsilon_2 d_1^2}{\varepsilon_1} + \frac{E_2}{d_1}\frac{\varepsilon_2 d_1^2}{\varepsilon_1} - V'_{GS}\right\} \tag{27}$$

For region III ($L_2 \leq x \leq L_3$)

$$\frac{d^2\varphi_{s3}(x)}{dx^2} - \frac{\varepsilon_1}{\varepsilon_2 d_1^2}\varphi_{s3}(x) = \frac{\varepsilon_1}{\varepsilon_2 d_1^2}\left\{\frac{-qN_D}{\varepsilon_2}\frac{\varepsilon_2 d_1^2}{\varepsilon_1} + \frac{E_3}{d_1}\frac{\varepsilon_2 d_1^2}{\varepsilon_1} - V'_{GS}\right\} \tag{28}$$

The solutions of equations (26) – (28) are obtained using the sum of complementary function (CF) and particular integral (PI) and given as follows

$$\varphi_{s1}(x) = Ae^{k_1 x} + Be^{-k_1 x} - p_1 \quad (29)$$

$$\varphi_{s2}(x) = Ce^{k_2 x} + De^{-k_2 x} - p_2 \quad (30)$$

$$\varphi_{s3}(x) = Ee^{k_3 x} + Fe^{-k_3 x} - p_3 \quad (31)$$

Where

$$k_1^2 = k_2^2 = \frac{\varepsilon_1}{\varepsilon_2 d_1^2} \quad (32)$$

$$p_1 = p_3 = \frac{-qN_D}{\varepsilon_2 k_1^2} + \frac{E_1}{d_1 k_1^2} - V'_{GS} \quad (33)$$

$$p_2 = \frac{-qN_D}{\varepsilon_2 k_2^2} + \frac{E_2}{d_1 k_2^2} - V'_{GS} \quad (34)$$

Now the constants of (29) – (31) are calculated using boundary conditions given in (13), (14) and (19) – (22), and given as

$$A = \frac{e^{-k_1 L_3}}{e^{-k_1 L_3} - e^{k_1 L_3}} \left( V_{bi} + p_1 + \frac{p_2 - p_1}{2e^{-k_1 L_1}} + \frac{p_3 - p_2}{2e^{-k_1 L_2}} \right)$$
$$- \frac{e^{k_1 L_3}}{e^{-k_1 L_3} - e^{k_1 L_3}} \left( \frac{p_1 - p_2}{2e^{k_1 L_1}} + \frac{p_2 - p_3}{2e^{k_1 L_2}} \right) - \left( \frac{V_{bi} + V_{DS} + p_3}{e^{-k_1 L_3} - e^{k_1 L_3}} \right) \quad (35)$$

$$B = \frac{e^{k_1 L_3}}{e^{k_1 L_3} - e^{-k_1 L_3}} (V_{bi} + p_1) - \frac{e^{-k_1 L_3}}{e^{-k_1 L_3} - e^{k_1 L_3}} \left( \frac{p_2 - p_1}{2e^{-k_1 L_1}} + \frac{p_3 - p_2}{2e^{-k_1 L_2}} \right)$$
$$- \frac{e^{k_1 L_3}}{e^{-k_1 L_3} - e^{k_1 L_3}} \left( \frac{p_1 - p_2}{2e^{k_1 L_1}} + \frac{p_2 - p_3}{2e^{k_1 L_2}} \right) - \left( \frac{V_{bi} + V_{DS} + p_3}{e^{-k_1 L_3} - e^{k_1 L_3}} \right) \quad (36)$$

$$C = \frac{e^{-k_1 L_3}}{e^{-k_1 L_3} - e^{k_1 L_3}} \left( V_{bi} + p_1 + \frac{p_2 - p_1}{2e^{-k_1 L_1}} + \frac{p_3 - p_2}{2e^{-k_1 L_2}} \right) - \frac{e^{-k_1 L_3}}{e^{-k_1 L_3} - e^{k_1 L_3}} \left( \frac{p_1 - p_2}{2e^{k_1 L_1}} \right)$$
$$- \frac{e^{k_1 L_3}}{e^{-k_1 L_3} - e^{k_1 L_3}} \left( \frac{p_2 - p_3}{2e^{k_1 L_2}} \right) - \left( \frac{V_{bi} + V_{DS} + p_3}{e^{-k_1 L_3} - e^{k_1 L_3}} \right) \quad (37)$$

$$D = \frac{e^{k_1 L_3}}{e^{k_1 L_3} - e^{-k_1 L_3}} (V_{bi} + p_1) - \frac{e^{k_1 L_3}}{e^{-k_1 L_3} - e^{k_1 L_3}} \left( \frac{p_2 - p_1}{2e^{-k_1 L_1}} \right)$$
$$- \frac{e^{-k_1 L_3}}{e^{-k_1 L_3} - e^{k_1 L_3}} \left( \frac{p_3 - p_2}{2e^{-k_1 L_2}} \right) - \frac{e^{k_1 L_3}}{e^{-k_1 L_3} - e^{k_1 L_3}} \left( \frac{p_1 - p_2}{2e^{k_1 L_1}} + \frac{p_2 - p_3}{2e^{k_1 L_2}} \right) \quad (38)$$
$$- \left( \frac{V_{bi} + V_{DS} + p_3}{e^{-k_1 L_3} - e^{k_1 L_3}} \right)$$

$$E = \frac{e^{-k_1 L_3}}{e^{-k_1 L_3} - e^{k_1 L_3}} \left( V_{bi} + p_1 + \frac{p_2 - p_1}{2e^{-k_1 L_1}} + \frac{p_3 - p_2}{2e^{-k_1 L_2}} \right) - \frac{e^{-k_1 L_3}}{e^{-k_1 L_3} - e^{k_1 L_3}} \left( \frac{p_1 - p_2}{2e^{k_1 L_1}} \right)$$
$$- \frac{e^{-k_1 L_3}}{e^{-k_1 L_3} - e^{k_1 L_3}} \left( \frac{p_2 - p_3}{2e^{k_1 L_2}} \right) - \left( \frac{V_{bi} + V_{DS} + p_3}{e^{-k_1 L_3} - e^{k_1 L_3}} \right) \quad (39)$$

$$F = \frac{e^{k_1 L_3}}{e^{k_1 L_3} - e^{-k_1 L_3}} (V_{bi} + p_1) - \frac{e^{k_1 L_3}}{e^{-k_1 L_3} - e^{k_1 L_3}} \left(\frac{p_2 - p_1}{2e^{-k_1 L_1}}\right)$$
$$- \frac{e^{k_1 L_3}}{e^{-k_1 L_3} - e^{k_1 L_3}} \left(\frac{p_3 - p_2}{2e^{-k_1 L_2}}\right) - \frac{e^{k_1 L_3}}{e^{-k_1 L_3} - e^{k_1 L_3}} \left(\frac{p_1 - p_2}{2e^{k_1 L_1}} + \frac{p_2 - p_3}{2e^{k_1 L_2}}\right) \quad (40)$$
$$- \left(\frac{V_{bi} + V_{DS} + p_3}{e^{-k_1 L_3} - e^{k_1 L_3}}\right)$$

## 3. Analysis of Surface Potential, Channel Potential, Electric Field, and Drain Current

Now the surface potential for the entire channel length ($x = 0$ to $L_3$) using (29), (30), (31) is given as

$$\varphi_s(x) = \varphi_{s1}(x) + \varphi_{s2}(x) + \varphi_{s3}(x) \quad (41)$$

Furthermore, using (23), (24), and (25) and substituting $y = d_1$, the channel potential for entire channel length ($x = 0$ to $L_3$) is given as

$$\varphi(x, d_1) = \varphi_1(x, d_1) + \varphi_2(x, d_1) + \varphi_3(x, d_1) \quad (42)$$

where channel potential for the region I, II, and III are given as

$$\varphi_1(x, d_1) = \varphi_{s1}(x) + \frac{\varepsilon_1}{\varepsilon_2}(\varphi_{s1}(x) - V'_{GS}) + \frac{-E_1}{2} \quad (43)$$

$$\varphi_2(x, d_1) = \varphi_{s2}(x) + \frac{\varepsilon_1}{\varepsilon_2}(\varphi_{s2}(x) - V'_{GS}) + \frac{-E_2}{2} \quad (44)$$

$$\varphi_3(x, d_1) = \varphi_{s3}(x) + \frac{\varepsilon_1}{\varepsilon_2}(\varphi_{s3}(x) - V'_{GS}) + \frac{-E_3}{2} \quad (45)$$

The lateral electric field along the 2DEG channel can be defined as

$$E_x = -\frac{d\varphi(x)}{dx} \quad (46)$$

And from $x = 0$ to $L_3$, the $x$ component of the electric field is given as

$$E_x = E_{x1} + E_{x2} + E_{x3} \quad (47)$$

Where lateral electric field components in the region I, II, III are obtained using (42), (43), (44) and (45) given as

$$E_{x1} = -\frac{d\varphi_{s1}(x)}{dx}\left(1 + \frac{\varepsilon_1}{\varepsilon_2}\right) \quad (48)$$

$$E_{x2} = -\frac{d\varphi_{s2}(x)}{dx}\left(1 + \frac{\varepsilon_1}{\varepsilon_2}\right) \quad (49)$$

$$E_{x3} = -\frac{d\varphi_{s3}(x)}{dx}\left(1 + \frac{\varepsilon_1}{\varepsilon_2}\right) \quad (50)$$

The drain current density is given as follows

$$I_{DS}(x)/W = qn_s(x)v_e(x) \tag{51}$$

Where W is the gate width, $v_e(x)$ is the electron effective velocity, and is given as

$$v_e(x) = \frac{\mu_n E_x v_{sat}}{\mu_n E(x) + v_{sat}} \tag{52}$$

Where electron mobility $\mu_n$ is given as per the negative differential mobility model from [21]

$$\mu_n = \frac{\mu_{n0} + \frac{v_{sat}}{E_x}\left(\frac{E_x}{E_C}\right)^\gamma}{1 + \left(\frac{E_x}{E_C}\right)^\gamma} \tag{53}$$

For β-Ga₂O₃, the parameters $\mu_{n0}$ = 140 cm² /Vs, $v_{sat}$ = 1.5 × 10⁷ cm /s, critical field $E_C$ = 2.25 × 10⁵ V /cm, and γ = 2.84 are taken from [24]. After substituting $n_s$ from (17), the drain current given by (51) can be rewritten as

$$I_{DS}(x) = WC_G(V_{GO} - E_F)\frac{\mu_n E_x v_{sat}}{\mu_n E_x + v_{sat}} \tag{54}$$

## 4. Results and Discussion

The simulation results for the device structure (Fig. 1) are obtained using the 2D device simulator—ATLAS [21]. The physics-based models and material properties are discussed in previous sections. The device simulator is first calibrated with experimental results measured in [5]. The mathematical model data are obtained using the computational tool—MATLAB [25]. The surface potential profile, channel potential, and electric field profile are extracted from structure files saved at different bias voltages. The surface potential plots are shown in Fig. 2. These data are extracted from at the heterointerface of different structure files for three different $V_{GS}$ of 1 V, –1 V, and –3 V and constant $V_{DS}$ of 10 V. The computed data from derived model (41) are also shown.

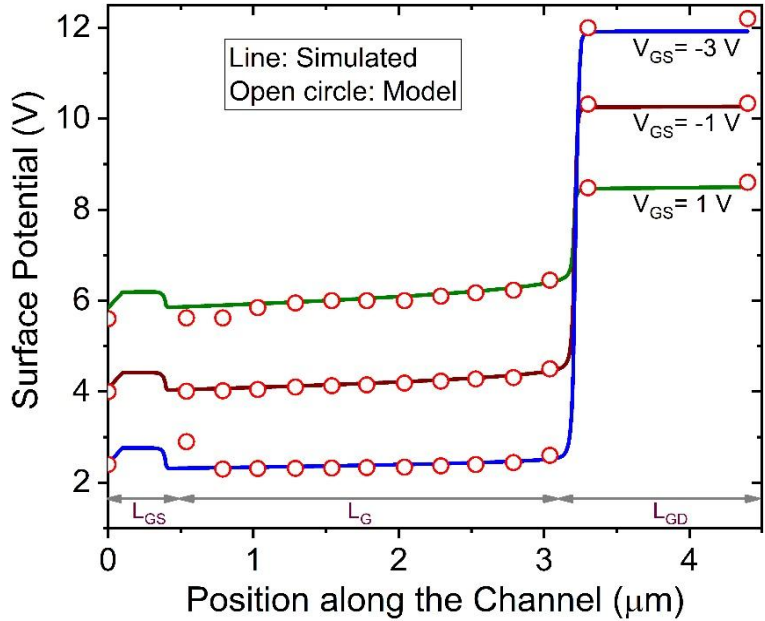

Fig. 2: The surface potential at the AlN/β-Ga₂O₃ interface for different $V_{GS}$ = 1, –1, –3 V at fixed $V_{DS}$ = 10 V. Full channel length subdivided into $L_{GS}$, $L_G$, and $L_{GD}$ is also drawn in the lower part of the figure.

The channel potential and electric field simulated data are extracted in the channel 5 nm below the AlN/β-Ga$_2$O$_3$ interface. The model data for channel potential and electric field are computed from (42), and (47) respectively. Simulated and model data for channel potential and electric field are shown in Fig. 3, and Fig. 4 respectively.

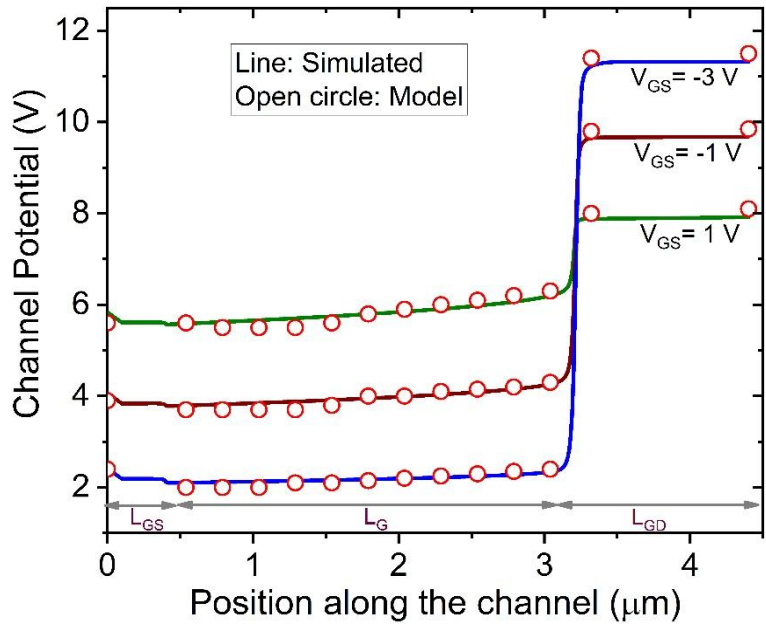

Fig. 3: Simulated and modeled data of channel potential along the channel below 5 nm from heterointerface for different $V_{GS}$ = 1, −1, −3 V at fixed $V_{DS}$ = 10 V. The plot is almost similar to surface potential plot except under $L_{GS}$ in the latter.

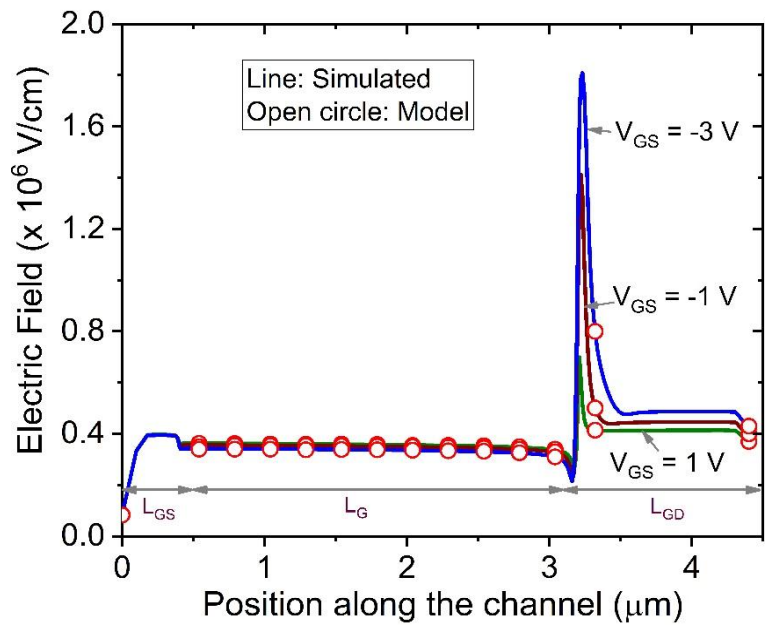

Fig. 4: Lateral electric field along the channel below 5 nm from heterointerface for different $V_{GS}$ = 1, −1, −3 V at fixed $V_{DS}$ = 10 V. Peak electric field of $1.8 \times 10^6$ V/cm is evident under the drain side gate edge.

The transfer characteristic of the proposed device is shown in Fig. 5. Drain current model data and experimental result relating to similar dimension device is also indicated. Except for the sub-threshold region, there is a good consistency between model data and simulation results. Additionally, a fair degree of agreement is evident with experimental data reported in [5].

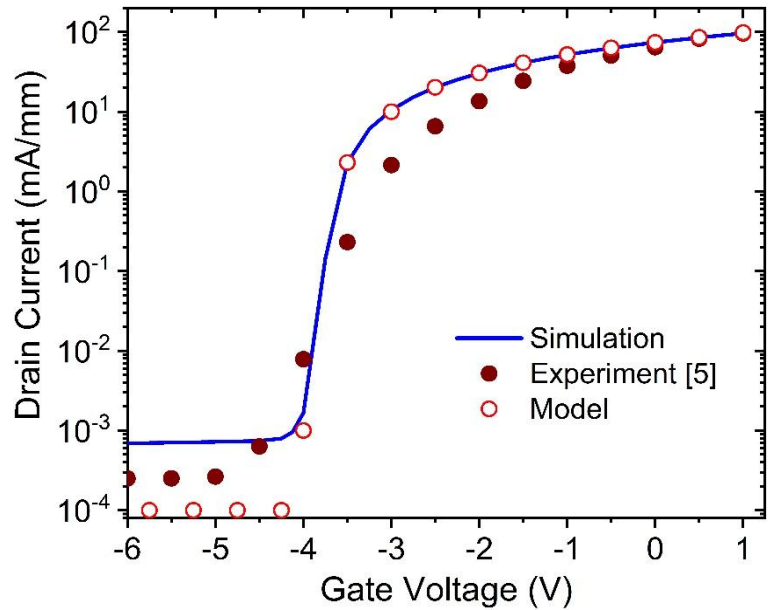

Fig. 5: Simulated, Experimental, and modeled data of $I_D - V_G$ characteristics of the analysed device.

## 5. Conclusion

The analytical model of surface potential, channel potential, and electric field for AlN/β-Ga$_2$O$_3$ is presented. The 2D Poisson equation is solved to develop an expression of potential and lateral field along the channel and consequently extended to find expression of drain current. For the validation of the simulation and model data, the simulation deck is calibrated with device dimensions similar to the experimental one as reported elsewhere. Analytical model data of surface potential, channel potential, and electric field are rigorously validated with different values of gate voltages. Furthermore, drain current model results are compared with simulation results as well as validated by experimental data. A good level of agreement among all three is obtained. Further work in the physics-based analytical models for the β-Ga$_2$O$_3$-based devices can be helpful to understand device mechanism and useful to predict its potential for futuristic high power applications.